\documentclass[english,aps,pop,reprint,superscriptaddress,showpacs,floatfix]{revtex4-1}
\usepackage{graphicx}
\usepackage{subfigure}
\usepackage{float}
\usepackage{calrsfs} 
\usepackage{dcolumn}
\usepackage{bm}
\usepackage{color}
\usepackage[normalem]{ulem}
\usepackage{subfigure}
\usepackage{sidecap}
\usepackage{wrapfig}
\usepackage{amsmath}
\usepackage{amssymb}
\usepackage{babel}
\usepackage{amsmath}
\usepackage{ulem}
\usepackage{pifont}
\usepackage{multirow}

\begin{document}

\begin{abstract}

It is shown that rapid substantial changes in heating rate can induce transitions to improved energy confinement regimes in zero-dimensional models for tokamak plasma phenomenology. We examine for the first time the effect of step changes in heating rate in the models of E-J.Kim and P.H.Diamond, \textit{Phys.Rev.Lett.} {\bf 90}, 185006 (2003) and M.A.Malkov and P.H.Diamond, \textit{Phys.Plasmas} {\bf 16}, 012504 (2009) which nonlinearly couple the evolving temperature gradient, micro-turbulence and a mesoscale flow; and in the extension of H.Zhu, S.C.Chapman and R.O.Dendy, \textit{Phys.Plasmas} {\bf 20}, 042302 (2013), which couples to a second mesoscale flow component. The temperature gradient rises, as does the confinement time defined by analogy with the fusion context, while micro-turbulence is suppressed. This outcome is robust against variation of heating rise time and against introduction of an additional variable into the model. It is also demonstrated that oscillating changes in heating rate can drive the level of micro-turbulence through a period-doubling path to chaos, where the amplitude of the oscillatory component of the heating rate is the control parameter.


\end{abstract}

\title{Transitions to improved confinement regimes induced by changes in heating in zero-dimensional models for tokamak plasmas}

\author{H.~Zhu}
\affiliation{Centre for Fusion, Space and Astrophysics, Department of Physics, University of Warwick, Coventry CV4 7AL, United Kingdom}
\author{S.C.~Chapman}
\affiliation{Centre for Fusion, Space and Astrophysics, Department of Physics, University of Warwick, Coventry CV4 7AL, United Kingdom}
\affiliation{Max Planck Institute for the Physics of Complex Systems, Dresden, Germany}
\affiliation{Department of Mathematics and Statistics, University of Tromso, Norway}
\author{R.O.~Dendy}
\affiliation{Euratom/CCFE Fusion Association, Culham Science Centre, Abingdon, Oxfordshire OX14 3DB, United Kingdom}
\affiliation{Centre for Fusion, Space and Astrophysics, Department of Physics, University of Warwick, Coventry CV4 7AL, United Kingdom}
\affiliation{Itoh Research Center for Plasma Turbulence, Kyushu University, Kasuga 816-8580, Japan}
\author{K.~Itoh}
\affiliation{National Institute for Fusion Science, Toki 509-5292, Japan}
\affiliation{Itoh Research Center for Plasma Turbulence, Kyushu University, Kasuga 816-8580, Japan}

\maketitle

\begin{flushleft}
\textbf{1. Introduction}
\end{flushleft}

Zero-dimensional models\cite{DLCT94, BG03, KD03, MD09, MD10, MD11, MD12, MD131, MD132, MD133, MD134, B10, II11, ZCD13, DMK13} --– that is, systems of coupled nonlinear differential equations with a single independent parametric coordinate representing time --– play an important role in interpreting fusion plasma behaviour. By choosing variables to represent key macroscopic quantities such as the temperature gradient $N$, the strength of micro-turbulence $E$, and the magnitude of large-scale coherent nonlinear structures $U$,  zero-dimensional models can be constructed in a manner that reflects the global phenomenology of, for example, L-mode and H-mode confinement physics. This enables empirically inspired physical models, which typically include predator-prey or Lotka-Volterra dynamics, to be tested and explored quantitatively: a necessary step, given that the dynamics can be    strongly nonlinear. It has not previously been established whether, in zero-dimensional models, rapid substantial increases in externally applied heating can engender sharp transitions in confinement properties, akin to heating-induced transition from L-mode to H-mode confinement in tokamak plasmas. First identified in Ref.\cite{W82}, the role of heating in this transition has been examined experimentally in all large tokamak plasmas, including DIII-D\cite{B94}, JT-60U\cite{F97}, Alcator C-Mod\cite{G97} and ASDEX-U\cite{R98}, and in a range of tritium, deuterium-tritium and hydrogen plasmas in JET\cite{R99}. For recent reviews of the H-mode and related fundamental plasma phenomena, see for example Refs\cite{W07, TFM09, DHM11}. Here we address heating-induced transitions in the framework of the well-established zero-dimensional model of Diamond, Kim and Malkov\cite{KD03,MD09}, hereafter KD/MD, which couples the three variables $(N,E,U)$ introduced above, and is driven by the external heating power $q(t)$, using the normalization of\citep{MD09}, see also\cite{ZCD13}:

\begin{eqnarray}
\frac{dE}{dt} &=& \left(N - N^4 - E - U \right) E \\
\frac{dU}{dt} &=&  \nu\left( \frac{E}{1 + \zeta N^4} - \eta\right)U \\
\frac{dN}{dt} &=& q\left(t\right) - \left( \rho + \sigma E\right)N  
\end{eqnarray} 
The meso-scale structures $U$ are induced by micro-turbulence $E$. The growth of micro-turbulence $E$ is suppressed and meso-scale structures $U$ as well as being self suppressed. External heating drives this system, and the external heating rate acts as a control parameter. The KD/MD model was recently extended in Ref.\cite{ZCD13}, hereafter ZCD, to include a fourth variable representing a second predator population $U_{2}$ of coherent nonlinear structures, for example geodesic acoustic modes(GAMs), in addition to the KD/MD population(denoted $U_{1}$ in ZCD) originally intended to represent zonal flows(ZFs). The introduction of these distinct classes of nonlinear structure in a zero-dimensional model follows the philosophy of Itoh \& Itoh\cite{II11}. We note that there is no direct interaction between $U_{1}$ and $U_{2}$ due to the parallelism of ZFs and GAMs\cite{DIIH05}.

The ZCD extension of the KD/MD model is written as\cite{ZCD13}:

\begin{eqnarray}
\frac{dE}{dt} &=& \left(N - N^4 - E - U_{1} - U_{2}\right) E\\
\frac{dU_{1}}{dt} &=&  \nu_{1} \left( \frac{E}{1 + \zeta N^4} - \eta_{1} \right)U_{1} \\
\frac{dU_{2}}{dt} &=&  \nu_{2} \left( \frac{E}{1 + \zeta N^4} - \eta_{2} \right)U_{2} \\
\frac{dN}{dt} &=& q\left(t\right) - \left( \rho + \sigma E\right)N
\end{eqnarray} 

In the present paper we investigate how sharp step changes in heating power $q(t)$ in the KD/MD model, and its ZCD extension, can induce confinement transitions. We also examine the impact of an oscillating heating rate on the ZCD extension, obtaining results for the system dynamics which differ significantly from those found in \cite{DMK13} for the KD/MD model. We quantify the dynamics, both in terms of the underlying $(N,E,U)$ phase space and in terms of the energy confinement time $\tau_{c}$ which is the key figure of merit. We identify the scaling of $\tau_{c}$ with heating power in the different confinement regimes of the KD/MD model and its ZCD extension. This is an important first step towards direct comparison between the global energy confinement times implicit in zero-dimensional models and the empirical confinement time scalings determined from multiple tokamak plasma experiments. We establish that the heating-induced confinement transitions are not strongly sensitive to the timescale on which heating power is increased. The results in the present paper are a significant step in the validation of the zero-dimensional approach, and of the KD/MD model and its ZCD extension, together with the physical identifications and assumptions which zero-dimensional models embody. 

A new result of \cite{ZCD13} concerned the attractive fixed point of the KD/MD model referred to as the transient mode (T-mode). It was found in ZCD that the introduction of a second coherent field $U_{2}$, acting as an additional predator on the micro-turbulence $E$, transforms the post-heating T-mode into a repulsive fixed point. A new attractive fixed point (see Fig.6 herein) or limit cycle (see Fig.7 herein) appears in ZCD, compare e.g. Fig.13 of ZCD. Here we refer to this new attractor as the oscillation mode (O-mode). Unlike the low confinement T-mode in KD/MD, the O-mode in ZCD has good confinement properties, as we discuss below. In the stable O-mode, $N$ is finite, as is $U_{2}$, with $E$ zero or very small, and $U_{1}$ zero. The present paper therefore also explores heating-induced transitions that can give rise to the O-mode in the ZCD extension of the KD/MD model.

We find that in the ZCD model, when the external heating rate includes a component that oscillates sinusoidally in time, as in \cite{DMK13}, a period-doubling path to chaos exists. The amplitude $A$ of the oscillatory component of the heating rate is the control parameter. The micro-turbulence level $E$ bifurcates with increasing $A$, and the ratio of values of $A$ at successive bifurcations is found to yield the first Feigenbaum constant\cite{F78} to high accuracy.

\begin{flushleft}
\textbf{2. Analytical confinement properties of the models}
\end{flushleft}

The energy confinement time in the KD/MD model and its ZCD extension can be addressed analytically, to some extent. We may define the energy confinement time $\tau_{c}$ at any instant by analogy with the fusion context\cite{F07}, using 

\begin{eqnarray}
\tau_{c} = \frac{N}{q - dN/dt}
\end{eqnarray} 
The structure of Eq.(8) is standard; in the present context, it reflects the fact that the temperature gradient $N$ is a physical proxy for stored energy, whose time evolution is driven by $q$ in Eq.(3). It follows from Eqs.(3), (7) and (8) that the confinement time in both KD/MD and ZCD is

\begin{eqnarray}
\tau_{c} = \frac{1}{\rho + \sigma E}
\end{eqnarray} 
for all $t$. At the fixed point, $dN/dt=0$ and 

\begin{eqnarray}
\tau_{F} = \frac{N_{F}}{q_{F}} = \frac{1}{\rho + \sigma E_{F}}
\end{eqnarray} 
where subscript F denotes evaluation at the fixed point.

We will need to solve numerically the system of equations Eqs.(1) to (3), and their ZCD counterparts Eqs.(4) to (7), in order to establish whether fixed points are accessible and how transitions between them (induced by changes in heating or otherwise) occur. However, provided the system can access the fixed point, Eq.(10) will hold and we can find $\tau_{F}$ for that fixed point. As discussed in KD/MD\cite{KD03,MD09}, the QH-mode fixed point has $E=U=0$. It therefore follows from Eq.(3) that in QH-mode there is linear scaling of stored energy with heating power,

\begin{eqnarray}
q_{QH} = \rho N_{QH}
\end{eqnarray} 
and from Eq.(10), $\tau_{QH}=1/\rho$. In contrast the T-mode fixed point\cite{MD09} has $E$ and $U$ finite, and Eq.(10) then yields 

\begin{eqnarray}
\tau_{T} = \frac{1}{\rho + \sigma \eta \left( 1 + \zeta N_{T}^{4} \right)} = \frac{\tau_{QH}}{1 + \left(\sigma \eta / \rho\right) \left( 1 + \zeta N_{T}^{4} \right)}
\end{eqnarray} 
This reflects the degradation of confinement in T-mode compared to QH-mode, associated with the level of T-mode turbulence $E_{T}=\eta \left( 1 + \zeta N_{T}^{4} \right)$. It further follows from Eq.(3) that 

\begin{eqnarray}
q_{T} = \rho N_{T}\left[1 + \left(\sigma \eta / \rho\right)\left( 1 + \zeta N_{T}^{4} \right)\right]
\end{eqnarray} 
in contrast to Eq.(11). Inversion of Eq.(13) to yield $N_{T}$ as a function of $q_{T}$ can be achieved numerically. Substitution of this result into Eq.(12) then yields $\tau_{T}$ as a function of $q_{T}$. In Fig.1 we plot the relative changes in the proxy for stored energy, $N_{QH}/N_{T}$, and in confinement time $\tau_{QH}/\tau_{T}$, as functions of the normalised increase in heating power $\delta q/q_{0}$ in the KD/MD model, where $\delta q=q_{QH}-q_{T}$ and $q_{0}=q_{T}$. The solid line in Fig.1 is derived from Eqs.(11) to (13), with over-plotted points derived from direct solution of Eqs.(1) to (3) and (8).

\begin{figure}
\begin{center}
\includegraphics[width=1\columnwidth]{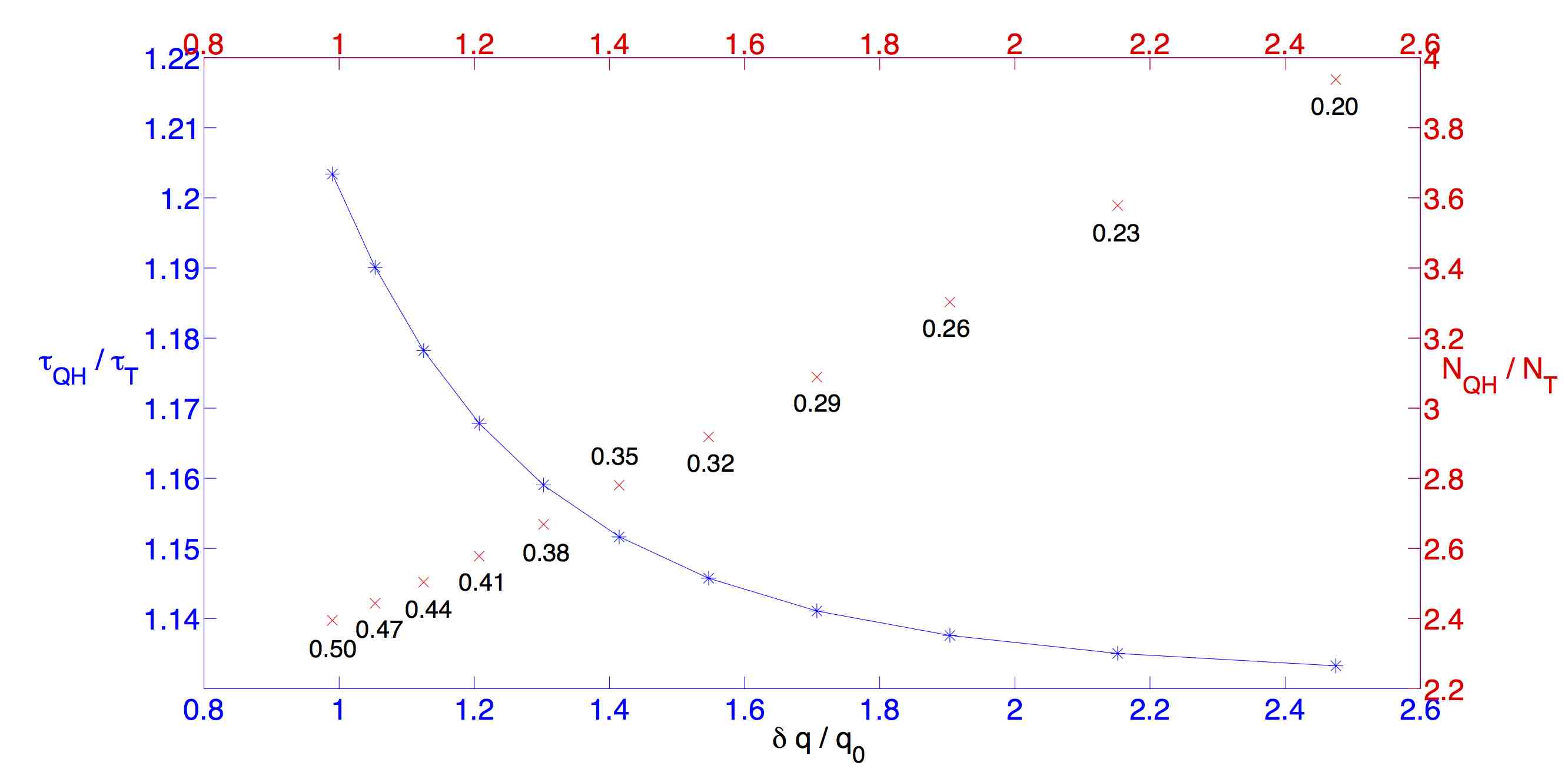}
\caption{KD/MD model dependence of ratios of confinement time $\tau_{QH}/\tau_{T}$ (blue stars; left scale) and temperature gradient $N_{QH}/N_{T}$ (red crosses; right scale) on the normalised increase in heating rate $\delta q/q_{0}$. Solid line for $\tau_{QH}/\tau_{T}$ is inferred from Eqs.(12) and (13). Points are obtained from numerical results for $\delta q=0.495$, $\tau_{QH}=1.8182$, $q_{0}$ values are shown in figure; other parameter values are $\nu = 19$, $\eta = 0.12$, $\rho = 0.55$, $\sigma = 0.6$, $\zeta = 1.7$.}
\end{center}
\end{figure}

The O-mode attractive fixed point or limit cycle of ZCD can be well approximated by $U_{1}=0$ and 

\begin{eqnarray}
E_{O} = \eta_{2} \left( 1 + \zeta N_{O}^{4} \right)
\end{eqnarray} 

\begin{eqnarray}
U_{2O} = N_{O} - N_{O}^{4} - E_{O}
\end{eqnarray} 
It follows that

\begin{eqnarray}
\tau_{O} = \frac{1}{\rho + \sigma \eta_{2} \left(1 + \zeta N_{O}^4 \right)}
\end{eqnarray} 
Analysis similar to that for the T-mode following Eq.(12) is then possible.

\begin{figure}
\begin{center}
\includegraphics[width=1\columnwidth]{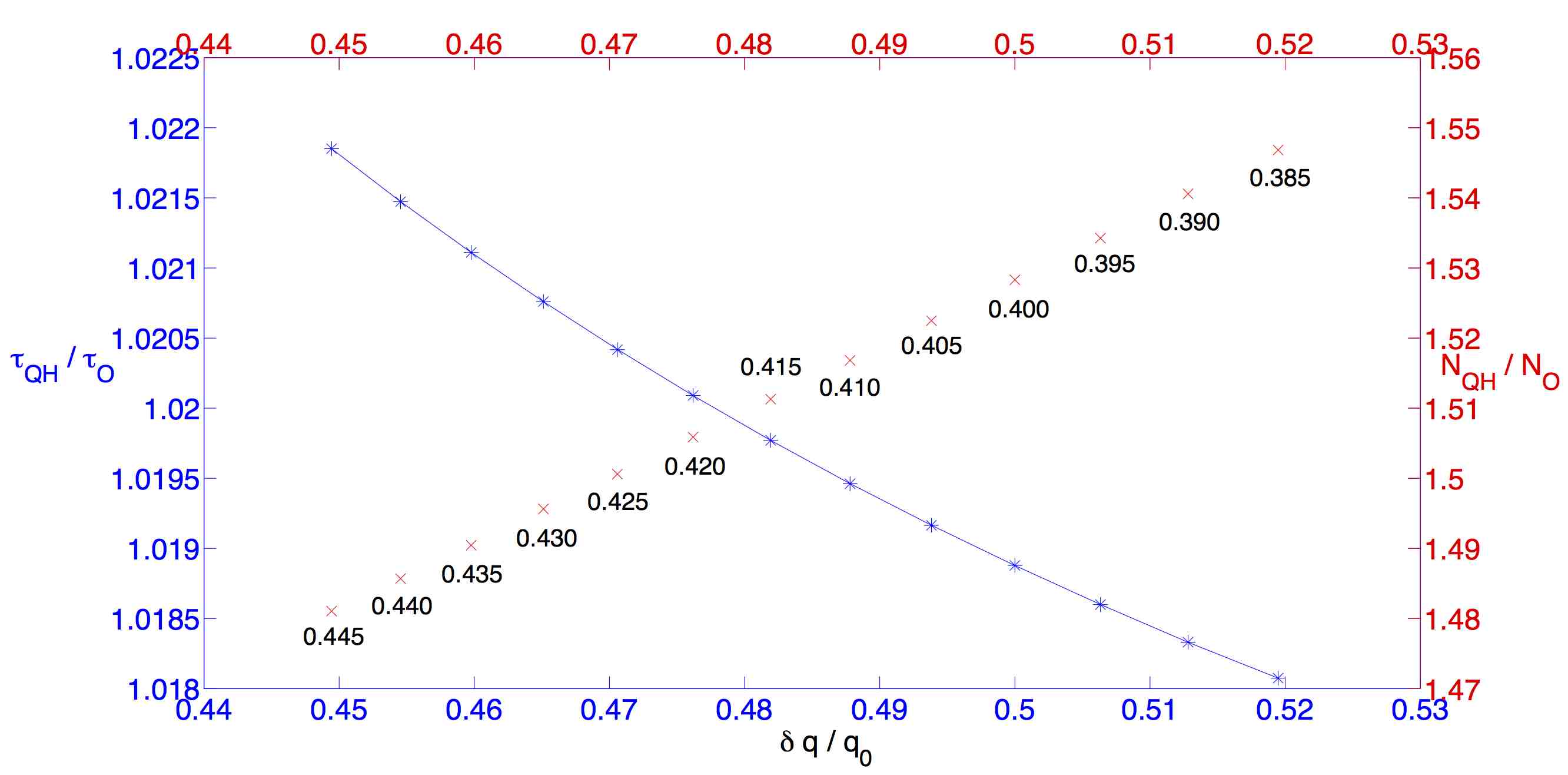}
\caption{ZCD model dependence of ratios of confinement time $\tau_{QH}/\tau_{O}$ (blue stars; left scale) and temperature gradient $N_{QH}/N_{O}$ (red crosses; right scale) on the normalised increase in heating rate $\delta q/q_{0}$. Solid line is inferred from Eqs.(11) and (16). Points are obtained from numerical results for $\delta q=0.20$, $\tau_{QH}=1.8182$, $q_{0}$ values are shown in figure; other parameter values are $\nu_{1} = 19$, $\nu_{2} = 0.19$, $\eta_{1} = 0.12$, $\eta_{2} = 0.012$, $\rho = 0.55$, $\sigma = 0.6$, $\zeta = 1.7$.}
\end{center}
\end{figure}

In the present paper, the T-mode is the lower confinement regime, compared to the enhanced confinement QH-mode in KD/MD and ZCD, and also compared to the O-mode in ZCD. The model QH-mode has highly idealised confinement properties embodied in Eq.(11). Figure 2 shows that the confinement properties of the ZCD post-heating O-mode are very similar to QH-mode, although weakly dependent on heating. These good confinement regimes effectively provide the benchmark with respect to which the degraded T-mode confinement is normalised. Figures.1 and 2 provide a general method to parametrise the energy confinement transition properties of zero-dimensional models in similar terms to experiments, see for example the classic studies of tokamak plasma confinement scaling in Ref.\cite{K97} for L-mode and Ref.\cite{I99} for H-mode.

\begin{flushleft}
\textbf{3. Confinement transitions induced by changes in heating in the KD/MD model}
\end{flushleft}

Understanding the confinement properties of the fixed point attractors and limit cycles of the KD/MD model and its ZCD extension is necessary, but not sufficient, for analysing the mapping from these zero-dimensional approaches to tokamak phenomenology. The transient time evolution of the system variables towards and between fixed points can, as we shall see, be of long duration and may relate to tokamak scenarios. In the present paper, we focus particularly on changes in time evolution that are consequent on rapid changes in heating power $q$.   

\begin{figure}
\begin{center}
\includegraphics[width=1\columnwidth]{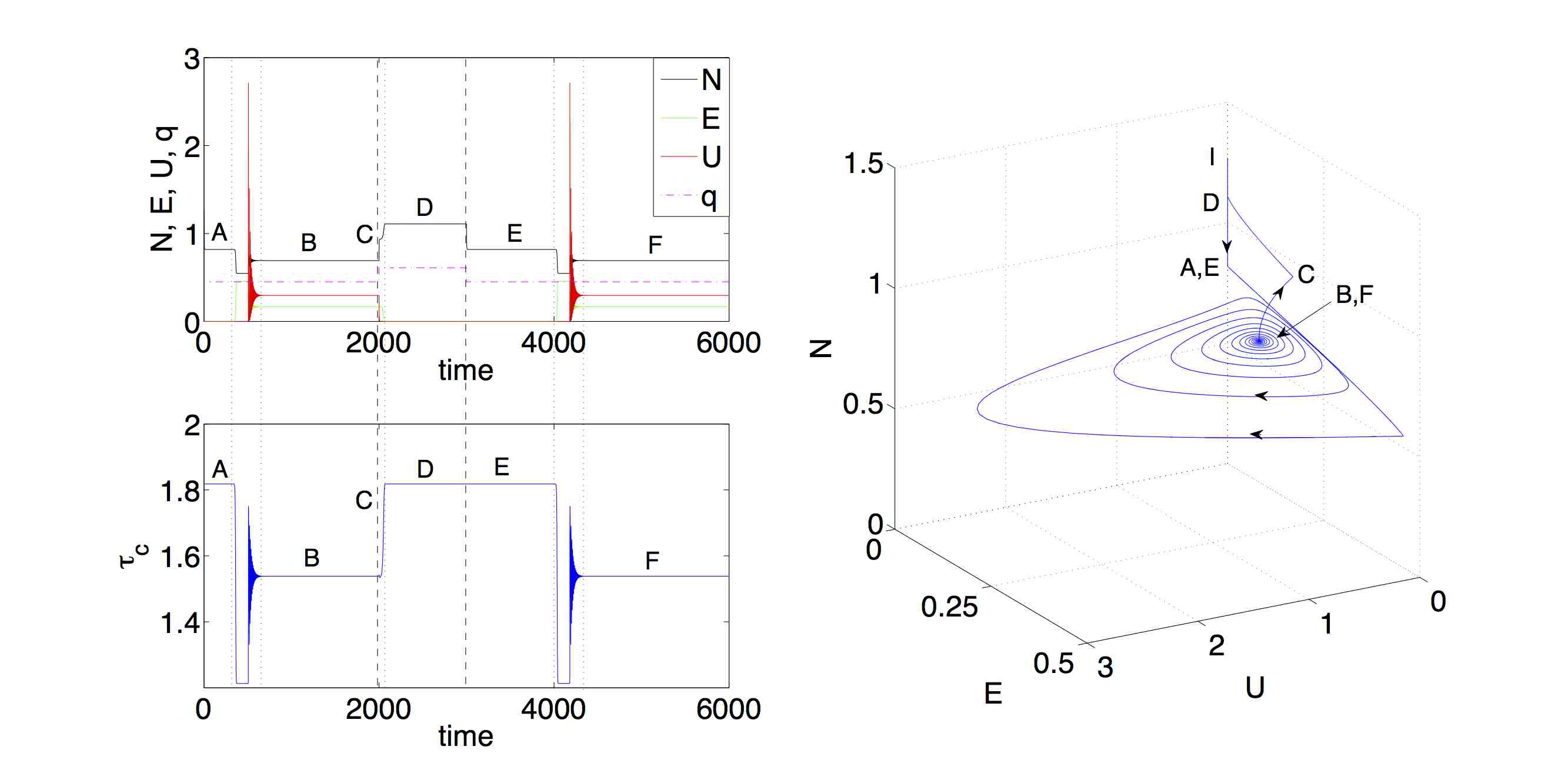}
\caption{Time traces and phase space evolution for the KD/MD model, Eqs(1) to (3), with a discontinuous increase in heating rate $q$ by amount $\delta {q} = 0.16$ from $q_{0} = 0.45$ at $t = 2000$; $q$ reverts to $q_{0}$ at $t = 3000$. Upper left plot shows time traces of variables $N$ (black), $q$ (dashed magenta), $U$ (red) and $E$ (green). Lower left plot shows time trace of energy confinement time $\tau_{c}$ defined by Eq.(8). Right plot shows time evolution of the system in $(N, E, U)$ phase space. The sequence of key phases is labelled in all three plots in this Figure as follows. A is the initial transient evolution from the over-powered H-mode point I, leading to convergent cyclic motion towards fixed point attractor B corresponding to T-mode. At C the instantaneous increase in heating rate $q$ induces rapid departure from the T-mode attractor B to the QH-mode (increased $N$; $E = U = 0$) attractor D with improved confinement time. Instantaneous reversion of $q$ to initial value $q_{0}$ brings the end of phase D and results in immediate transition to a QH-mode by exponential decrease in $N$, labelled E, with a lower value of $N$ and the same confinement time as phase D. There is later a spontaneous back transition from E at $t = 4000$, followed by convergent cyclic motion F to the T-mode attractor B.}
\end{center}
\end{figure}

Figures 3 to 5 display an example of the responses, in the KD/MD model, to a substantial instantaneous rise $\delta q$ in heating power $q$, which is then sustained at this higher level before later returning instantaneously to its initial level $q_{0}$. The resulting system dynamics --– a proxy for plasma phenomenology –-- is characterised in each Figure in terms of time traces of $N, E, U$ and $q$ (upper plot), and of $\tau_{c}$ (lower plot). The heating power $q_{0}$ is successively larger in the system shown in Fig.3 through Fig.5. Before $\delta q$ is applied, the system has relaxed to its attractor for $q_{0}$. For the particular parameter values chosen, in Figs.3 to 5 this fixed point is a state with relatively low $N$, and non-zero turbulence level $E$ and zonal flow amplitude $U$. In KD/MD, this low confinement fixed point is referred to as the transient mode (T-mode). 

Figure 3 shows that instantaneous application of $\delta q$ = 0.16 to the $q_{0}$ = 0.45 T-mode causes a transition to an improved confinement regime which is  identified by KD/MD with the quiescent H-mode (QH-mode). This is a fixed point which has larger $N$, while $E = U = 0$. As a consequence (from Eq.(11)), the value of $\tau_{c}$ rises instantly by about twenty per cent. At the termination of additional heating when $q \rightarrow q_{0}$, the system is still at this fixed point but now at lower $N$, with $E$ and $U$ still zero, hence still a QH-mode. The value of $\tau_{c}$ remains constant at its value for the additionally heated QH-mode, since $\tau_{QH}=1/ \rho$ is independent of $N$. This second QH-mode phase persists for some time after the heating power has reverted to the lower initial value $q_{0} = 0.45$. Eventually the system returns to the initial T-mode. 

\begin{figure}
\begin{center}
\includegraphics[width=1\columnwidth]{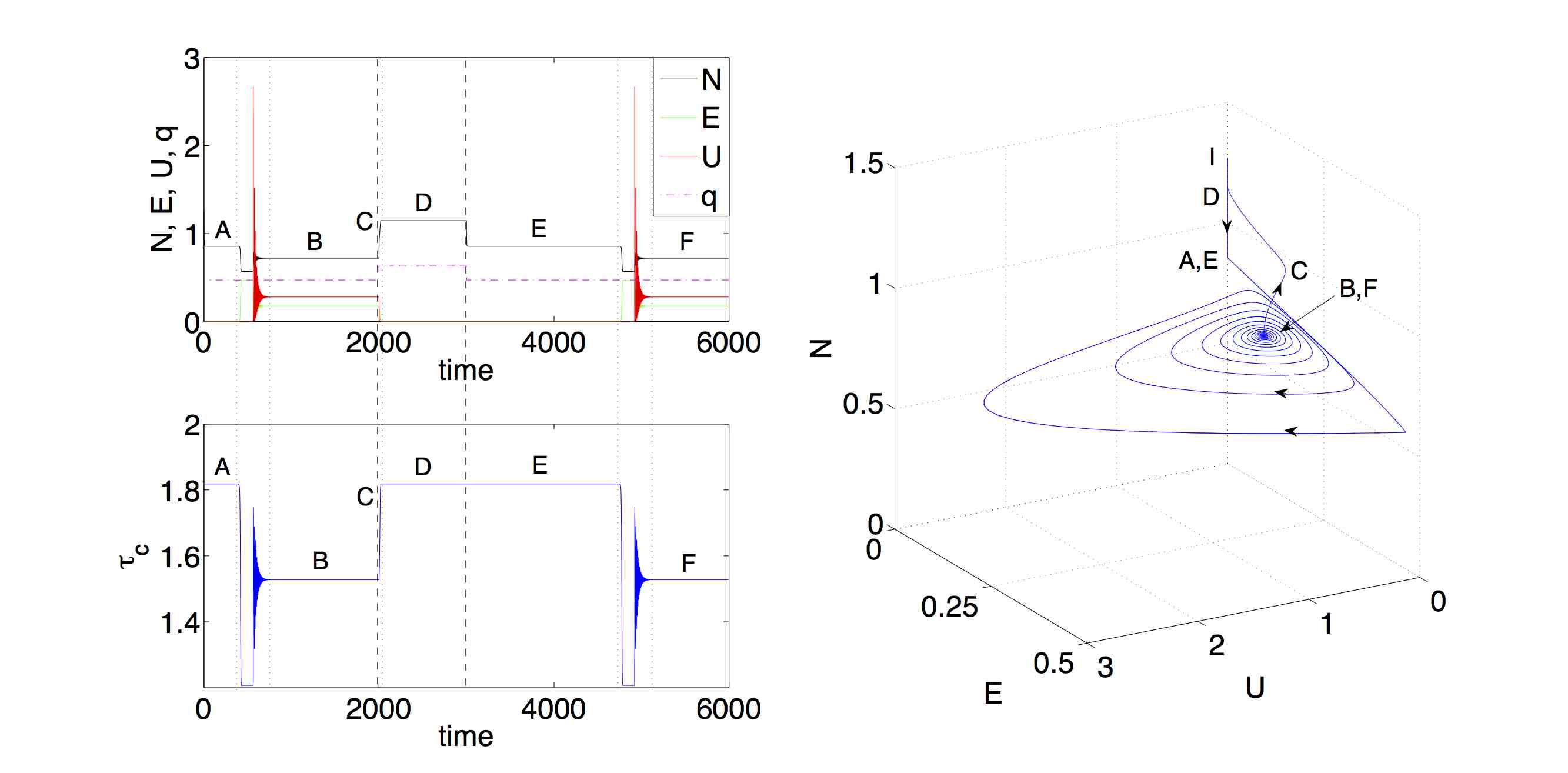}
\caption{As Fig.3, for the case where $q_{0} = 0.47$. The major difference is the longer duration of the post-heating QH-mode phase E, after reversion of $q$ to its initial value.}
\end{center}
\end{figure}

\begin{figure}
\begin{center}
\includegraphics[width=1\columnwidth]{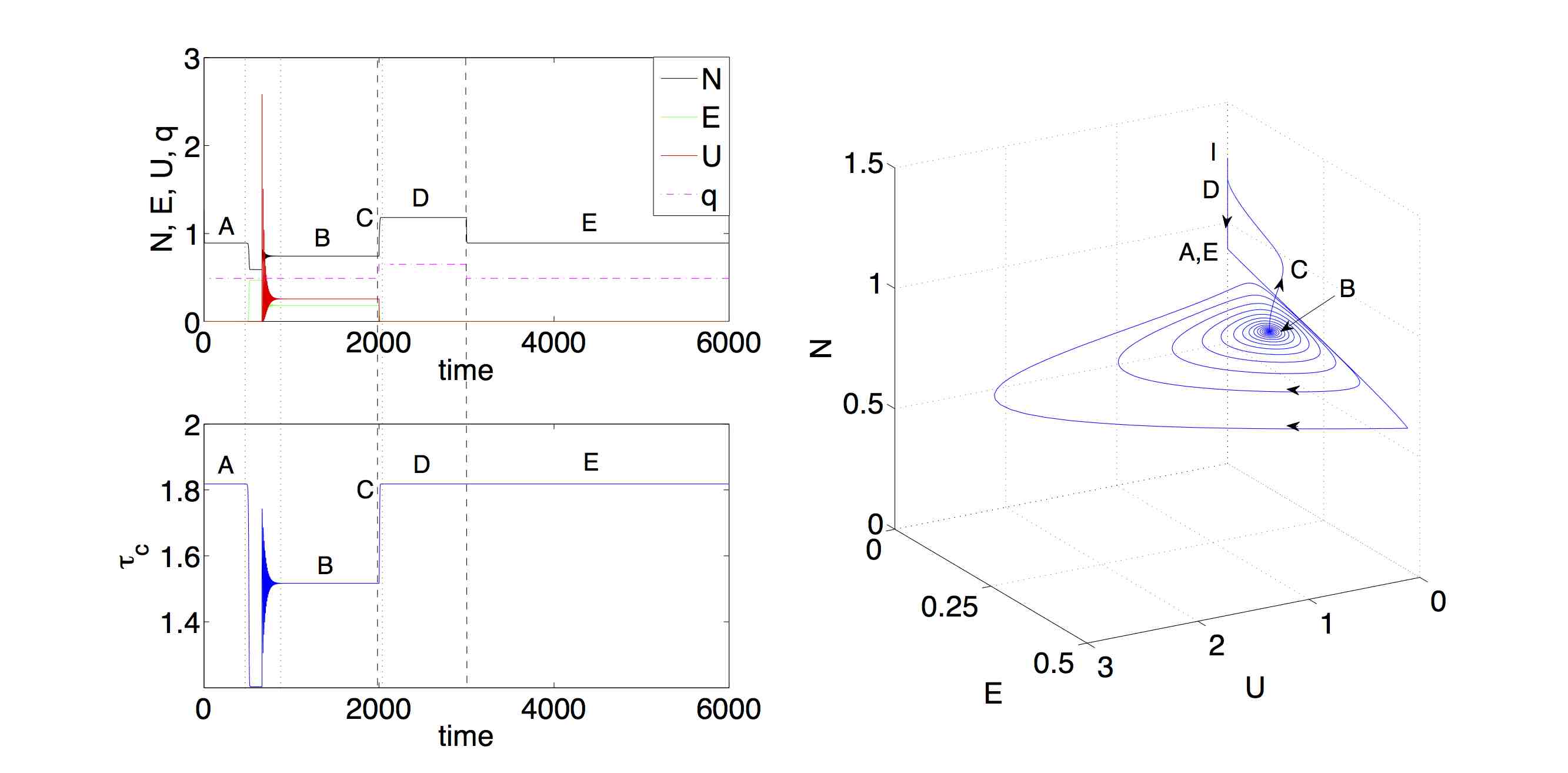}
\caption{As Fig.4, for the case where $q_{0} = 0.49$. The major difference is that the back transition from the post-heating QH-mode phase E, which is not a stable attractor, has not yet occurred by the end of this run.}
\end{center}
\end{figure}

In Figure 4, the duration of the QH-mode after the heating has reverted to its initial value $q_{0}$ is substantially longer than in Fig.3. Here the only parameter difference from Fig.3 is that $q_{0} = 0.47$, implying a slightly higher maximum power $q_{0} + \delta q$ and slightly lower fractional change $\delta q/q_{0}$. For the case shown in Fig.5, where $q_{0} = 0.49$, the system remains in the post-heating QH-mode until the run ends. An eventual back transition to T-mode after the heating power reverts to $q_{0} = 0.49$ has not had time to occur.

In conclusion, in Figs.3 to 5, the initial T-mode with confinement time $\tau_{T}$ is sustained by the heating rate $q_{0} = q_{T}$. The sharp rise in heating rate to $q_{0} + \delta q = q_{QH}$ triggers the transition to the QH-mode at higher $N$ and with improved confinement time $\tau_{QH}$.

\begin{flushleft}
\textbf{4. Confinement transitions induced by changes in heating in the ZCD model}
\end{flushleft}

\begin{figure}
\begin{center}
\includegraphics[width=1\columnwidth]{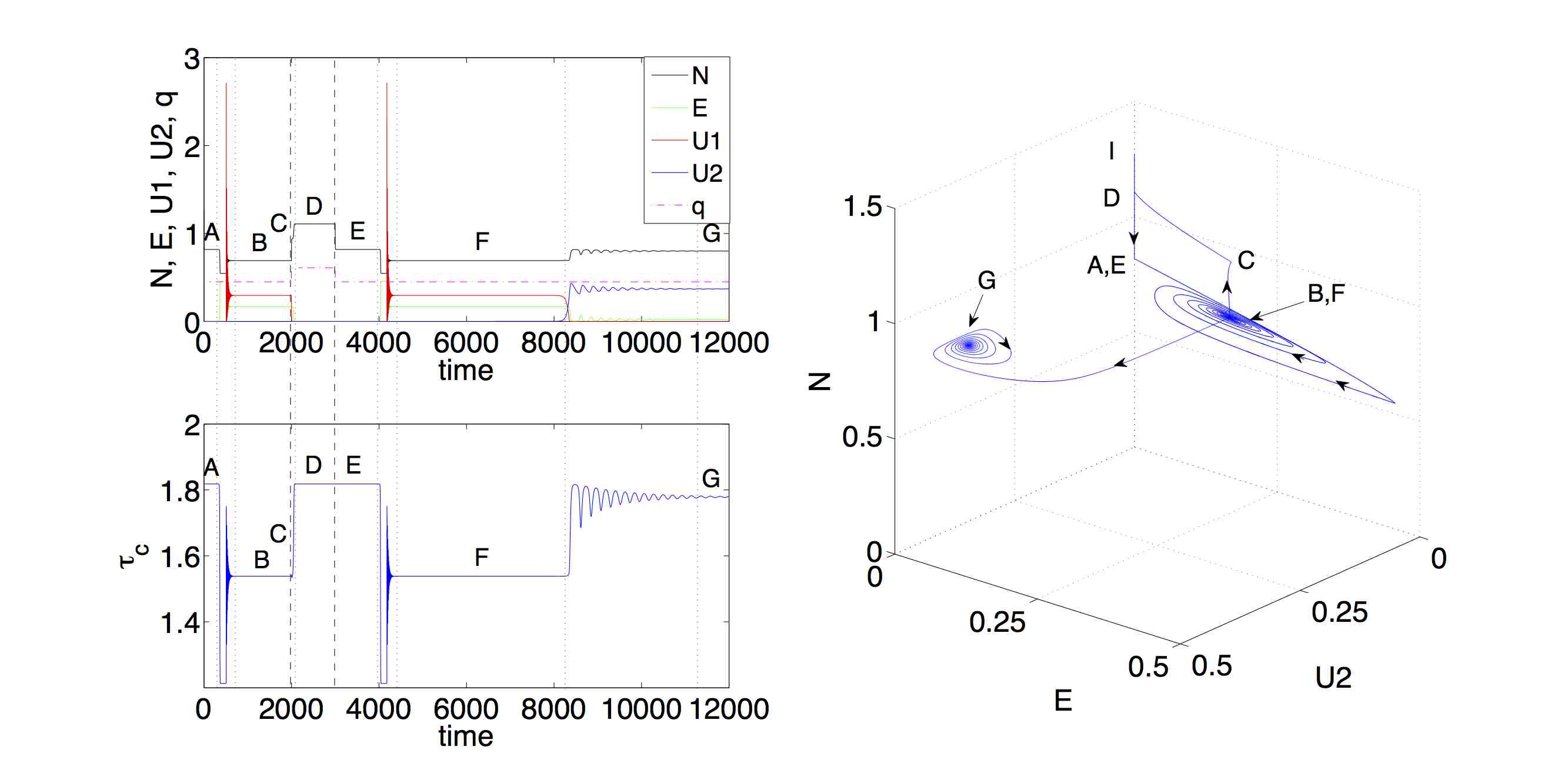}
\caption{As Fig.3, for the two-predator ZCD model, Eqs(4) to (7), with a sharp heating transition where $q_{0} = 0.45$, $\delta q = 0.16$. The second predator field $U_{2}$ is traced in blue in the upper left plot. The major difference from Fig.3 is that the post-heating T-mode state F is a repulsive fixed point, from which the system spontaneously transitions and converges cyclically to the fixed point G. This is known from Ref.\cite{ZCD13}) and has enhanced $N$ and finite $U_{2}$, with $E$ very small. Here we refer to the attractive fixed point G as an example of O-mode.}
\end{center}
\end{figure}

\begin{figure}
\begin{center}
\includegraphics[width=1\columnwidth]{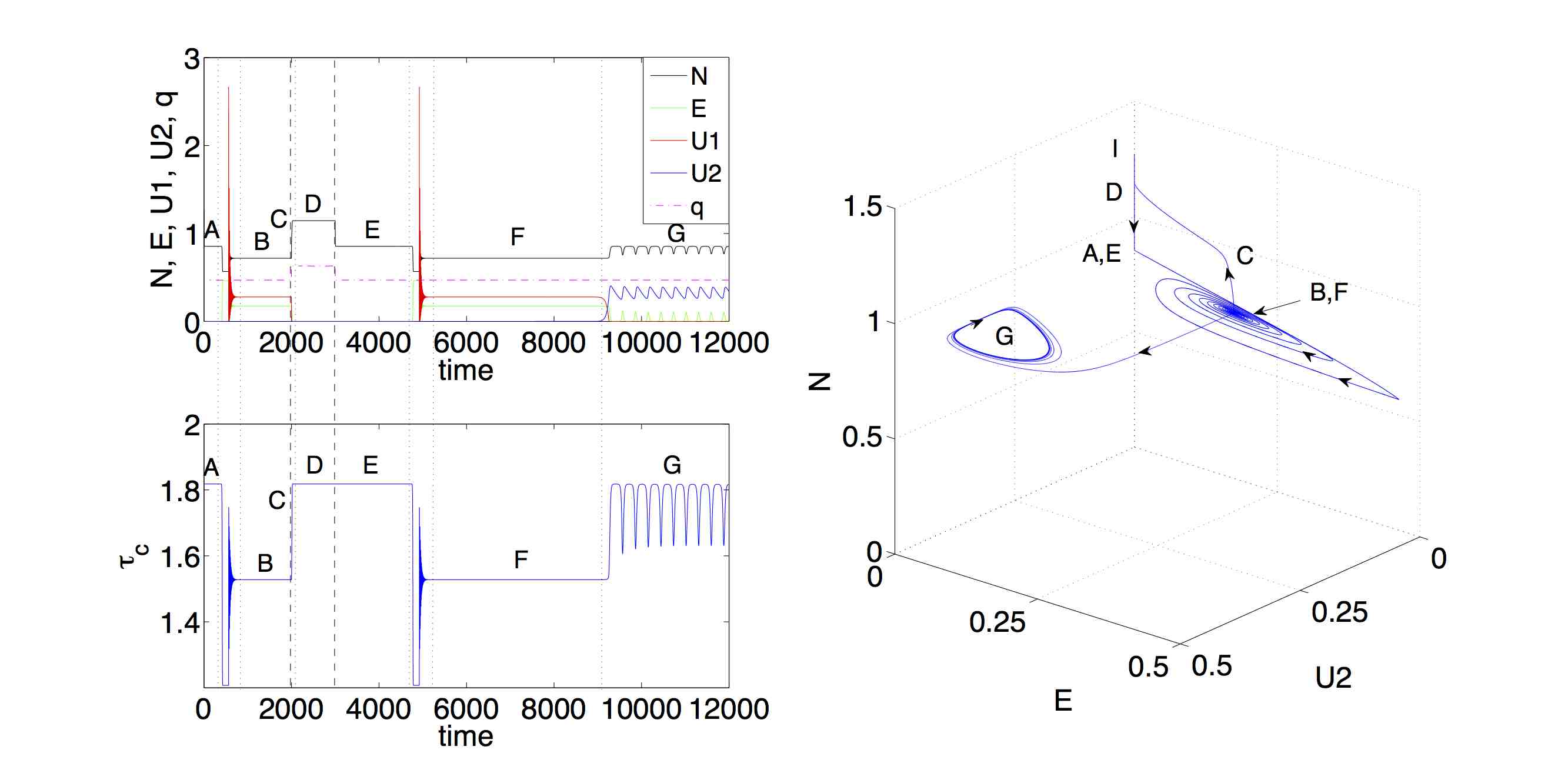}
\caption{As Fig.4, for the two-predator ZCD model\cite{ZCD13} with a sharp heating transition where $q_{0} = 0.47$, $\delta q = 0.16$. The major difference from Fig.4 is that the post-heating T-mode state F is a repulsive fixed point, from which the system spontaneously transitions and converges cyclically to the  limit cycle G. This is known from Ref.\cite{ZCD13} and has oscillations of enhanced $N$ and finite $U_{2}$, accompanied by small pulses of $E$. Here we refer to the attractive limit cycle G as an example of O-mode.}
\end{center}
\end{figure}

\begin{figure}
\begin{center}
\includegraphics[width=1\columnwidth]{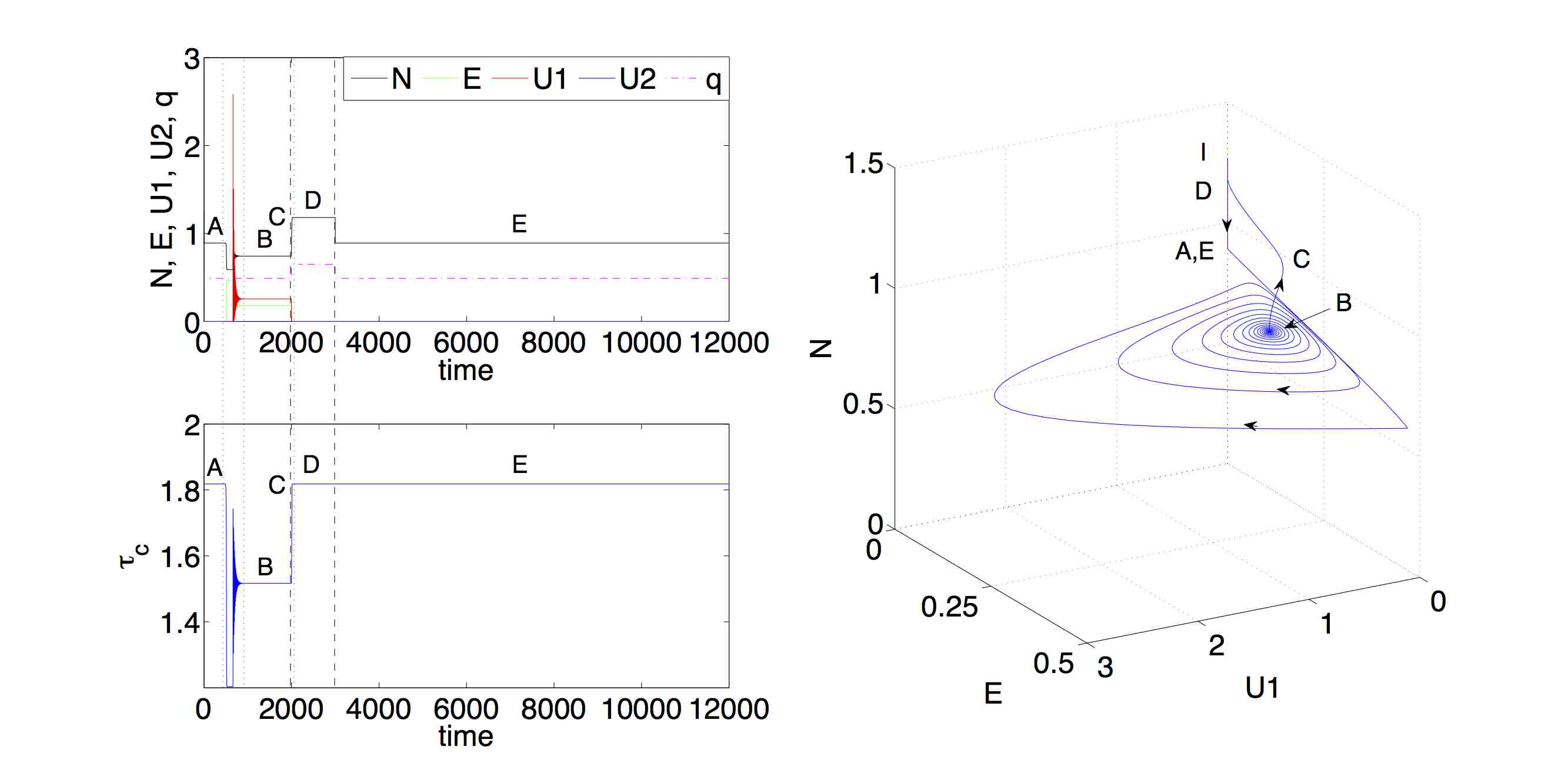}
\caption{As Fig.5, for the two-predator ZCD model\cite{ZCD13} with a sharp heating transition where $q_{0} = 0.49$, $\delta q = 0.16$. As in Fig.3, there is insufficient run time for the phase E QH-mode to transition to T-mode and then to the O-mode attractor, unlike Figs.6 and 7.}
\end{center}
\end{figure}

Figures 6 to 8 show the counterparts to Figs.3 to 5 that are obtained from the ZCD extension of the KD/MD model; that is, when a fourth variable $U_{2}$ representing a second coherent field predator is added to the KD/MD model, see Eqs.(4) to (7). The phenomenology of the heating-induced QH-mode and post-heating QH-mode in Figs.6 to 8 is very similar to that in the corresponding Figs.3 to 5 for the KD/MD model. From this we can infer that the induction of enhanced confinement by additional heating in the KD/MD model is robust against the introduction of a fourth variable as in ZCD. It is known\cite{ZCD13} that the eventual post-heating T-mode is a fixed point attractor in KD/MD but is a repulsive fixed point in ZCD. Figures 6 and 7 capture the transition from T-mode to the new ZCD fixed point (phase G in Fig.6) or limit cycle (phase G in Fig.7). This is the O-mode, with good confinement, as discussed in Sec.2.

\begin{flushleft}
\textbf{5. Impact of oscillating heating rate on the ZCD model}
\end{flushleft}

A topical question concerns the system response to oscillations, in time, of the heating rate about a constant value. The resultant changes in the phenomenology generated by zero-dimensional models are of interest both theoretically and, potentially, experimentally. Repeated on-off switching of electron cyclotron heating is now routine\cite{L01}, so that quasi-oscillatory ECH scenarios are becoming realisable. If one can identify distinctive signatures in the system response of a zero-dimensional model in such scenarios, this could assist a potential future experimental probe of the physical assumptions embodied in that model. We now examine the ZCD model in this context, and find that such a signature indeed exists, in the form of a classical period-doubling path\cite{H09} to chaos in the temperature gradient $N$, coherent field amplitude $U_{2}$, and level of micro-turbulence $E$, as the amplitude of the oscillatory component of the heating rate is increased. The response of the MD model, which has one fewer variable and does not appear to exhibit this distinctive phenomenology, was investigated in \cite{DMK13}.

We represent the heating rate by

\begin{eqnarray}
q\left(t\right) = q_{0} + Asin\left(\omega t\right)
\end{eqnarray} 
where $q_{0} = 0.47$, $\omega = 0.05$ and all other coefficients and initial conditions take the values that were used to generate Fig.7. This oscillatory timescale is fast compared to the duration of quasi-stationary phase in Figs.6 to 8. Specifically, the ratio of period of oscillating heating rate and that of limit cycle in Fig.7 is approximately 43.6 per cent. The control parameter in the following study is thus the amplitude $A$ of the oscillatory component of the heating rate defined in Eq.(17). Figures 9 to 11 show the initial period-doubling path from period-1, via period-2, to period-4 as the value of $A$ is increased from 0.0215 through 0.0240 to 0.0270. We note that these values of $A$ are correspond to a few per cent of the steady heating rate $q_{0} = 0.47$. Figures 9 to 11 all show: on the left, the power spectrum of $N$; on the right, the full attractor in $\left(N, U_2, E\right)$ space; and, inset, the time series of $N$. Figure 12 provides an additional perspective on this period-doubling by showing the power spectra of $N$ from Figs.9 to 11 over-plotted in the frequency range from 0.04 to 0.08. The fully chaotic attractor is shown in Fig.13, obtained for $A = 0.0295$. Figure 14 provides a comprehensive diagram of the period-doubling bifurcation path to chaos in the value of micro turbulence level $E$ as $A$ is increased from 0.0215 to 0.0295 in the ZCD model. We have obtained the values of $A_{n}$ at which the nth period-doubling bifurcations occur, from period-1 to period-8. We find $A_{1} = 0.0230, A_{2} = 0.0265, A_{3} = 0.0272$ and $A_{4} = 0.0273$, giving the ratios 4.666 which is within 0.05 per cent of the expected value 4.669\cite{F78}.

\begin{figure}
\begin{center}
\includegraphics[width=1\columnwidth]{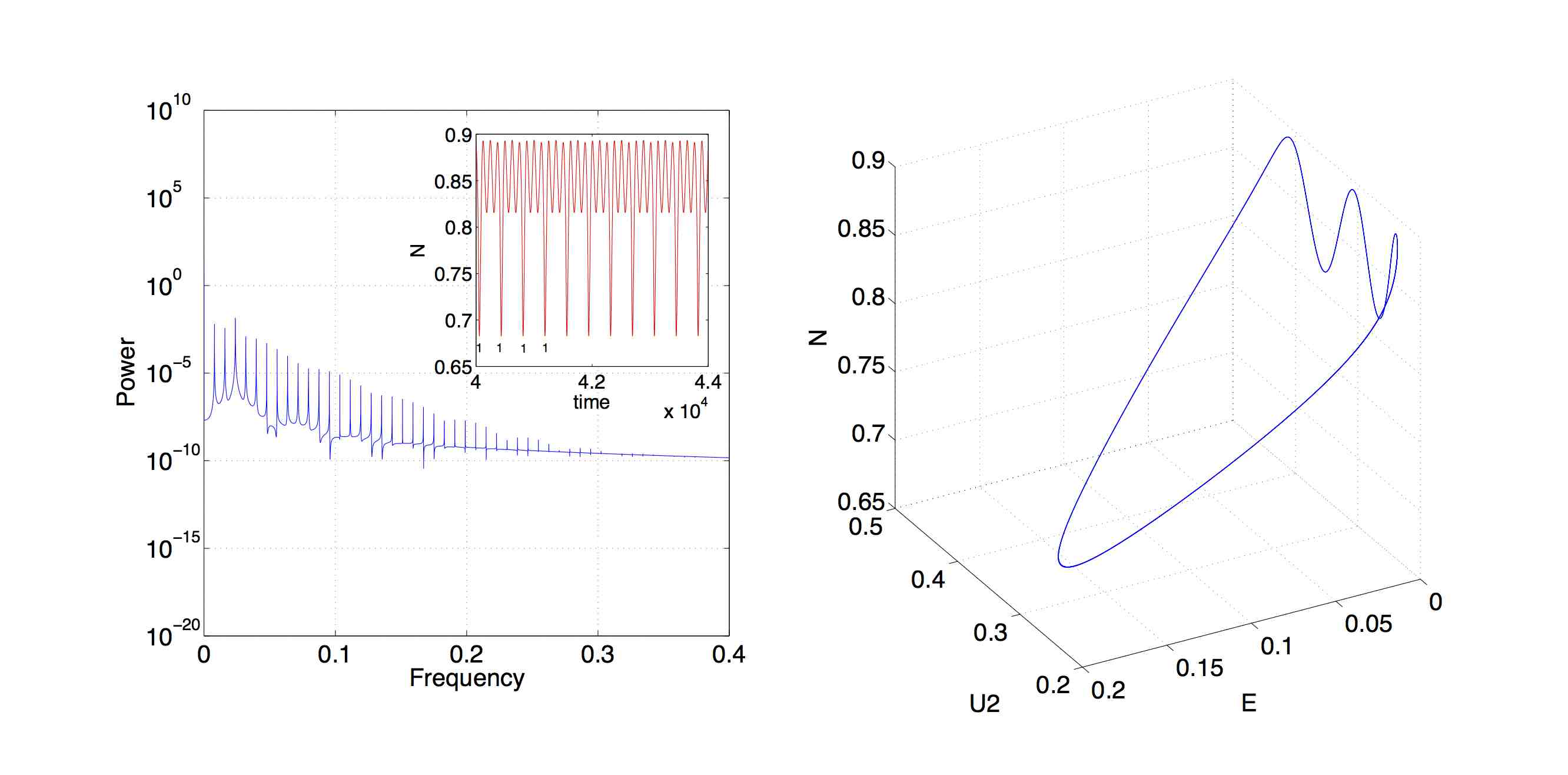}
\caption{Period-1 oscillation in ZCD system dynamics in response to the varying heating rate defined by Eq.(17) with $A = 0.0215$; other parameter values are as for Fig.7. Left, the power spectrum of $N$; right, the full attractor in $\left(N, U_2, E\right)$ space; inset, the time series of $N$.}
\end{center}
\end{figure}

\begin{figure}
\begin{center}
\includegraphics[width=1\columnwidth]{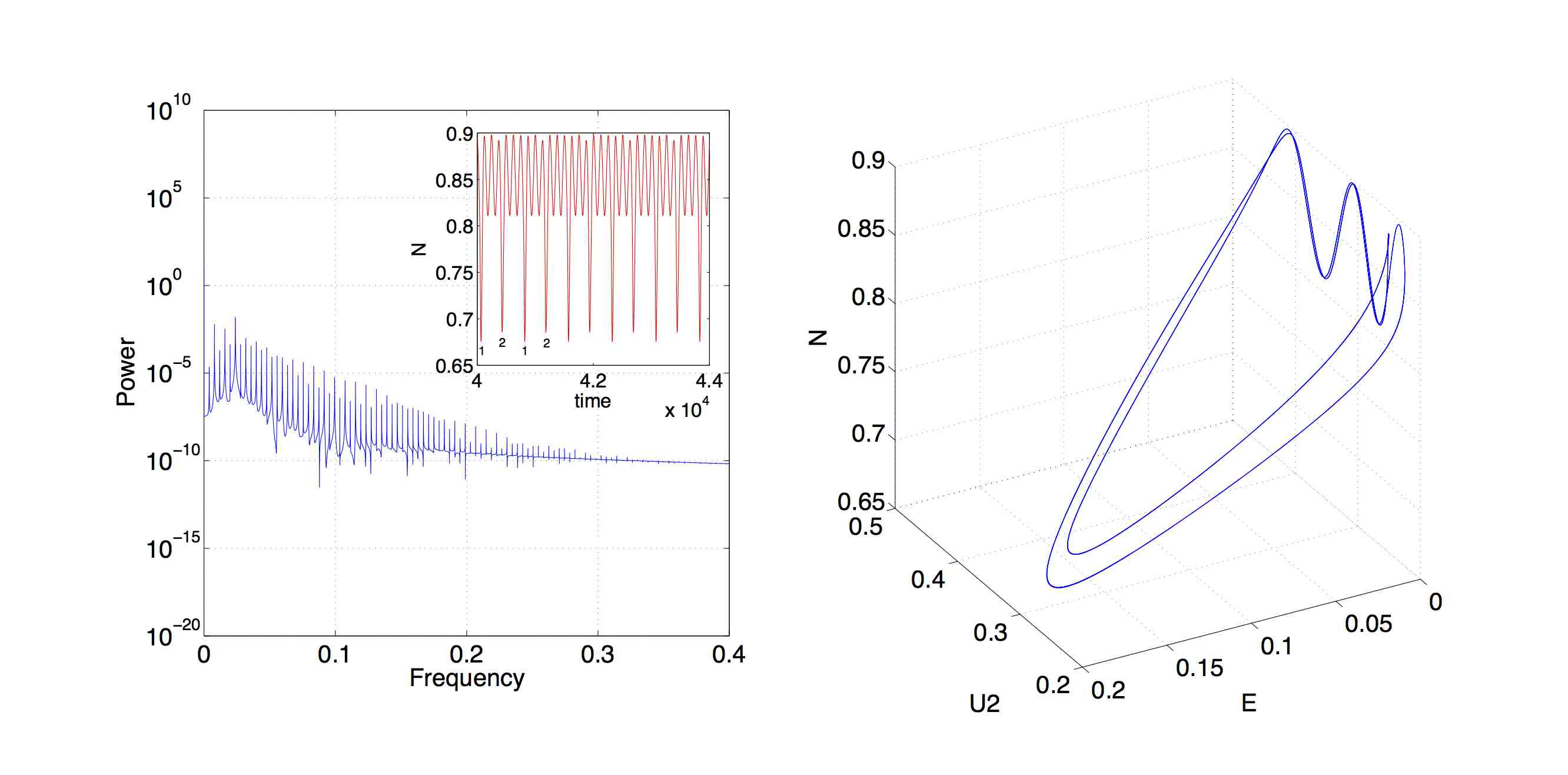}
\caption{As Fig.9, showing period-2 oscillation in ZCD system dynamics when $A = 0.0240$.}
\end{center}
\end{figure}

\begin{figure}
\begin{center}
\includegraphics[width=1\columnwidth]{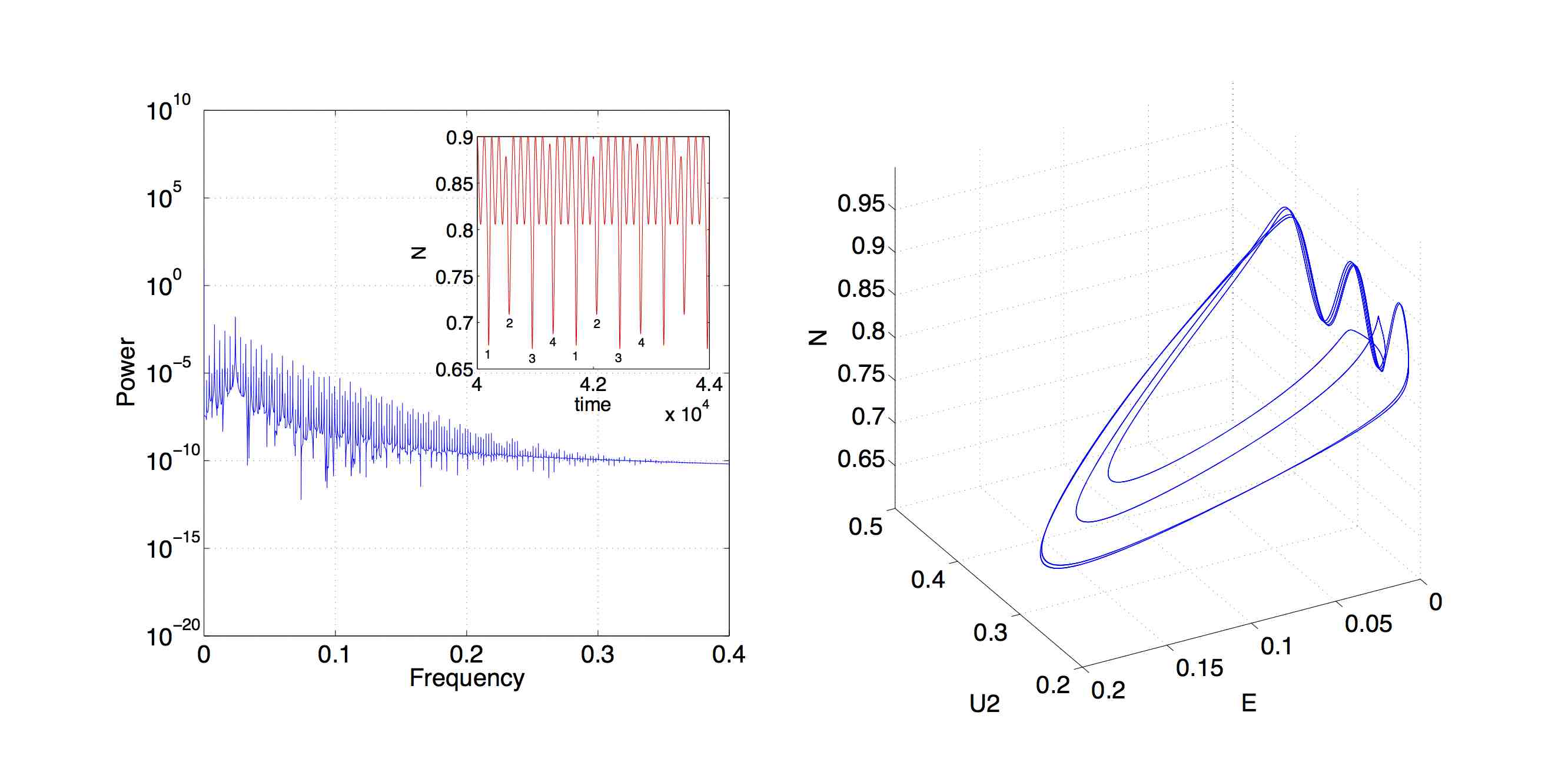}
\caption{As Fig.9, showing period-4 oscillation in ZCD system dynamics when $A = 0.0270$.}
\end{center}
\end{figure}

\begin{figure}
\begin{center}
\includegraphics[width=1\columnwidth]{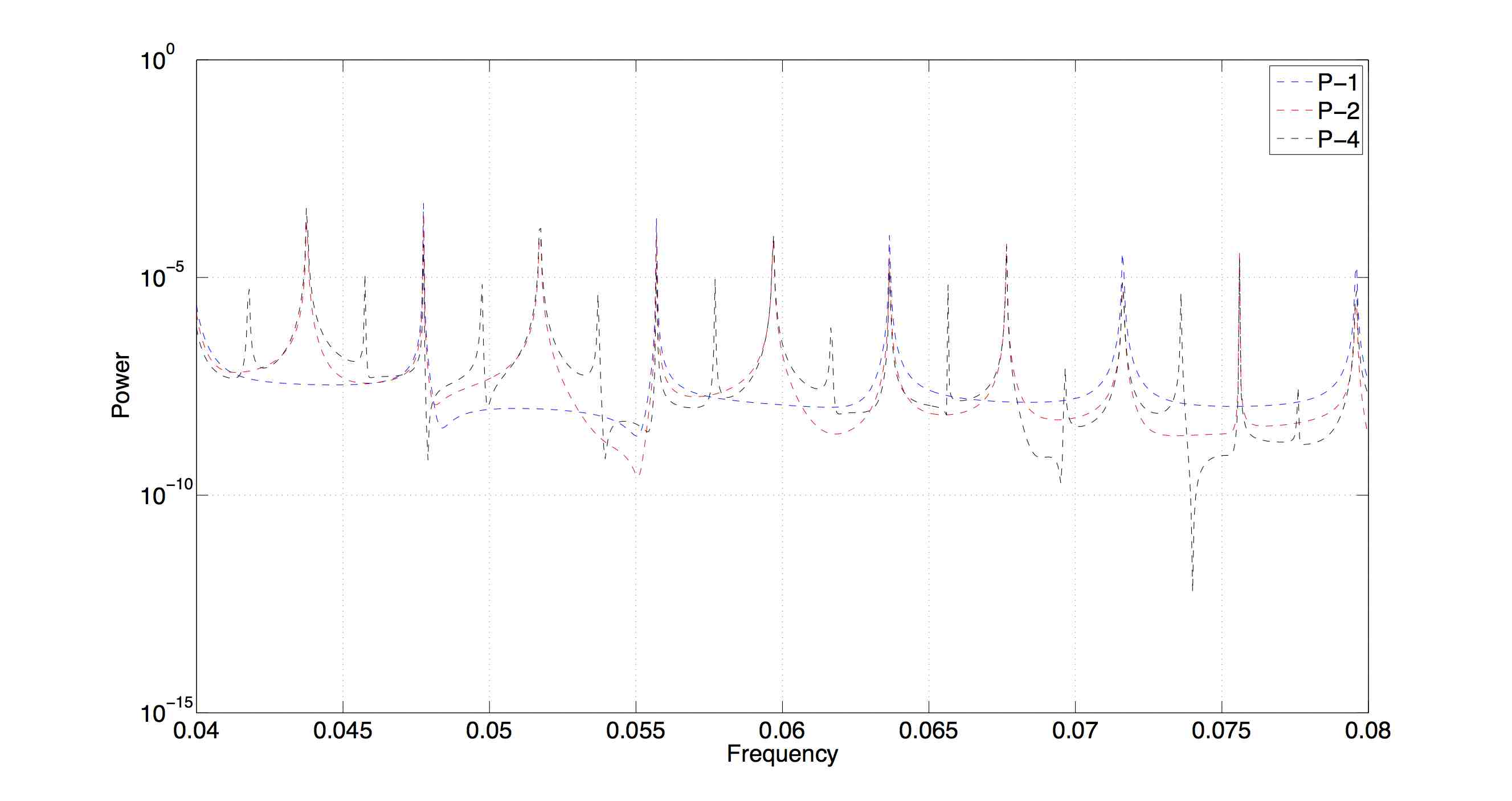}
\caption{Period-doubling illustrated by the power spectra of $N$ from Figs.9 to 11, over-plotted in the frequency range from 0.04 to 0.08. Blue, red and black dash lines denote spectra of period-1, period-2 and period-4 respectively.}
\end{center}
\end{figure}

\begin{figure}
\begin{center}
\includegraphics[width=1\columnwidth]{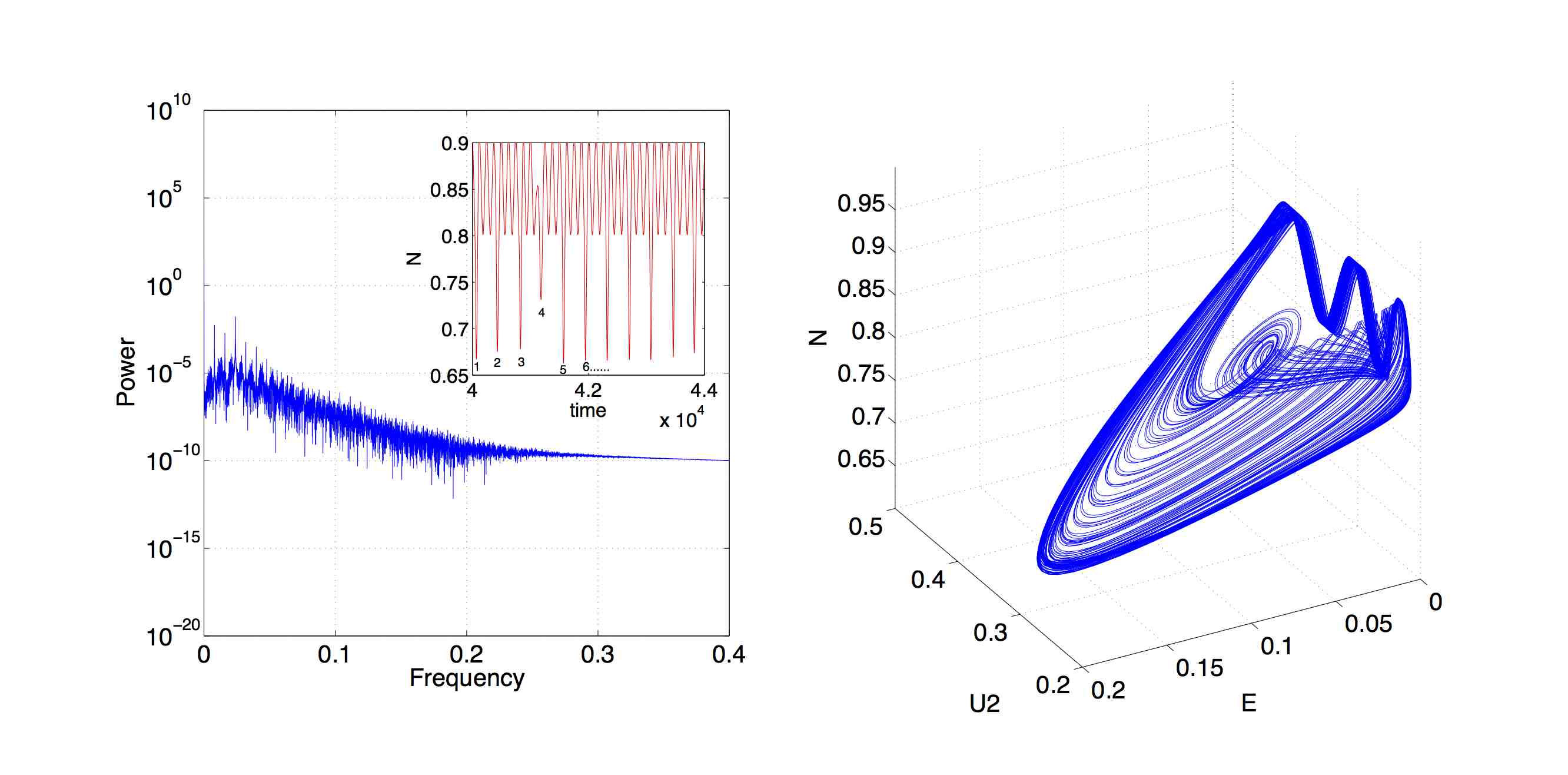}
\caption{As Fig.9, showing chaotic attractor of the ZCD system dynamics when $A = 0.0295$.}
\end{center}
\end{figure}

\begin{figure*}
\begin{center}
\includegraphics[width=2\columnwidth]{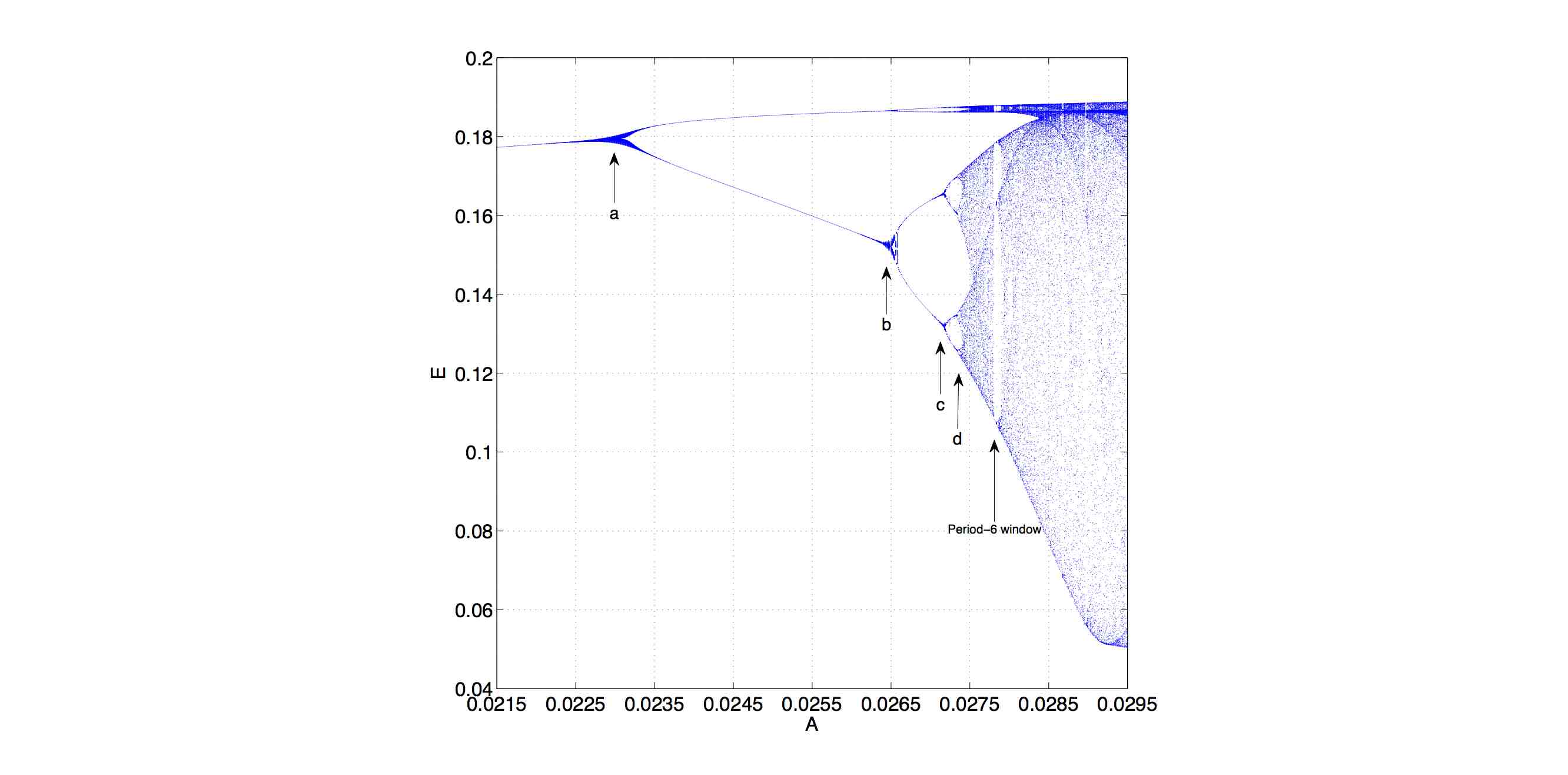}
\caption{Period-doubling bifurcation path to chaos of ZCD system dynamics. Micro-turbulence level $E$ is plotted versus amplitude $A$ of oscillatory heating component in Eq.(17), in the range 0.0215 to 0.0295. The first four arrows indicate successive bifurcations, which occur at values $A = 0.0230, 0.0265, 0.0272$ and $0.0273$. These yield Feigenbaum's ratio to within 0.05 per cent. The fifth arrow marks a period-6 window within the chaotic region.}
\end{center}
\end{figure*}

\begin{flushleft}
\textbf{6. Conclusion}
\end{flushleft}

In this paper we have shown that the KD/MD model and its ZCD extension capture a key feature of tokamak plasma confinement phenomenology, additional to those previously noted in Refs.\cite{KD03,MD09} and \cite{ZCD13}. Specifically, a rapid substantial change in heating power can trigger a transition to an enhanced confinement regime having steeper temperature gradient $N$, longer energy confinement time $\tau_{c}$, and suppressed micro-turbulence level $E$. Enhanced confinement is sustained throughout the duration of the heating pulse. It can continue after heating reverts to its initial level, retaining the same value of $\tau_{c}$ but with lower $N$, for a time whose duration depends on the values of the initial and additional levels of heating. Importantly, this tokamak plasma-like enhanced confinement phenomenology is robust: both against minor variations of the switch-on time scale for the additional heating, and against inclusion of a second predator field in the model. The latter step also creates a new attractive fixed point or limit cycle which has enhanced confinement characteristics. This O-mode has higher values of $N$ and $\tau_{c}$ than the lower confinement T-mode which precedes it, with nonlinear structure amplitudes $U_{1} = 0$ and $U_{2}$ finite, and micro-turbulence level small or zero. The KD/MD model and its ZCD extension also possess well defined scaling relations between energy confinement time and heating power, which can be calculated. We emphasise again that numerical solution of the time evolving system, as well as knowledge of its fixed points, are necessary for these studies.

From a dynamical systems perspective, we have identified a confinement time parameter $\tau_{c}$ which depends only on the values of the macroscopic fields $(E, U, N, q)$. The value of $\tau_{c}$ at the fixed points can be used to characterise the confinement states of the model. Since these values can be found analytically for any zero-dimensional model, we have provided a procedure to obtain the dependence of confinement time on the heating enhancement $\delta q/q_{0}$. If these fixed points are attractors in the model, then the duration of these confinement states will be long and will be insensitive to the detailed time dependence of heating $q(t)$. If on the other hand these fixed points are repulsive, then full numerical solution of the given zero-dimensional model equations is required to determine the duration of the corresponding confinement states, and whether this is sensitive to the detailed time dependence of the heating $q(t)$. 

When a small oscillatory-in-time component is added to the steady heating rate, we find that the ZCD model can exhibit a classic period-doubling path to chaos in, for example, the level of micro-turbulence $E$ as the amplitude of oscillation is increased. In this, the ZCD model may differ from the MD model, for which oscillatory heating was studied in \cite{DMK13}. This distinctive phenomenology may offer a path to future experimental testing of the assumptions of zero-dimensional models, and perhaps distinguishing between them.

We infer that the heating-induced transitions between confinement regimes are not strongly sensitive to the temporal sharpness of the change in heating rate from $q_{0}$ to $q_{0}+\delta q$. We have repeated the numerical experiment of Sections 3 to 4 for the same parameters, except that the heating transition up and down is now a continuous tanh function. We conclude that discontinuous and slightly smoothed changes of heating rate with time produce essentially similar results. Resilience against noise fluctuations in the heating has also been investigated. We find that the results are effectively invariant against noise in the heating at levels of 1\% to 10\%.

The results in this paper reinforce the apparent validity of the conceptually simple (albeit strongly nonlinear) physical picture embodied in the KD/MD model. Very few simple first principles models can spontaneously generate tokamak-like enhanced confinement phenomenology; the sandpile of Refs.\cite{CDR99,CDH01} is an example. It is increasingly clear that the KD/MD model is in this category. 

\begin{flushleft}
\textbf{Acknowledgements}
\end{flushleft}

We are grateful for helpful discussions with Professors S. Inagaki and S.-I. Itoh.
This work was part-funded by the support by KAKENHI (21224014, 23244113) from JSPS and the EPSRC and the RCUK Energy Programme under grant EP/I501045 and the European
Communities under the contract of Association between EURATOM and CCFE. The views and opinions expressed
herein do not necessarily reflect those of the European Commission.

\end{document}